\documentclass[journal=ancham,layout=onecolumn,manuscript=article]{achemso}

\usepackage{algorithm,algorithmic}

\usepackage{graphicx} 
\usepackage[colorlinks=False]{hyperref} 
\usepackage{amsmath}  
\usepackage{amsfonts} %
\usepackage{amssymb}  %
\usepackage{color}

\usepackage[version=3]{mhchem}
\graphicspath{{./figs/}}

{}

{

\author{Martin Robinson}
\email{martin.robinson@cs.ox.ac.uk}
\affiliation{Department of Computer Science, University of Oxford, Wolfson
Building, Parks Road, Oxford, OX1 3QD, United Kingdom.}

\author{Alexandr N Simonov}
\affiliation{School of Chemistry, Monash University, Clayton, Vic. 3800,
Australia.}

\author{Jie Zhang}
\affiliation{School of Chemistry, Monash University, Clayton, Vic. 3800,
Australia.}

\author{Alan Bond}
\email{alan.bond@monash.edu.au}
\affiliation{School of Chemistry, Monash University, Clayton, Vic. 3800,
Australia.}

\author{David Gavaghan}
\email{david.gavaghan@cs.ox.ac.uk}
\affiliation{Department of Computer Science, University of Oxford, Wolfson
Building, Parks Road, Oxford, OX1 3QD, United Kingdom.}

\title{Separating the effects of experimental noise from inherent system
variability in voltammetry: the $[$Fe(CN)$_6]^{3-/ 4-}$ process}

\begin{document}

\begin{abstract}

Recently, we have introduced the use of techniques
drawn from Bayesian statistics to recover kinetic and thermodynamic parameters
from voltammetric data, and were able to show that the technique of large
amplitude ac voltammetry yielded significantly more accurate parameter values
than the equivalent dc approach. In this paper we build on this work to show
that this approach allows us, for the first time, to separate the effects of
random experimental noise and inherent system variability in voltammetric
experiments. We analyse ten repeated experimental data sets for the
$[$Fe(CN)$_6]^{3-/ 4-}$ process, again using large-amplitude ac cyclic
voltammetry. In each of the ten cases we are able to obtain an extremely good
fit to the experimental data and obtain very narrow distributions of the
recovered parameters governing both the faradaic (the reversible formal
faradaic potential, $E_0$, the standard heterogeneous charge transfer rate
  constant
$k_0$, and the charge transfer coefficient $\alpha$) and non-faradaic terms
(uncompensated resistance, $R_u$, and double layer capacitance, $C_{dl}$). We
    then employ hierarchical Bayesian methods to recover the underlying
    ``hyperdistribution" of
the faradaic and non-faradaic parameters, showing that in general the variation
    between the experimental data sets is significantly greater than suggested
    by individual experiments, except for $\alpha$ where the
    inter-experiment variation was relatively minor. Correlations between pairs
    of parameters are provided, and for
    example, reveal a weak link between $k_0$ and $C_{dl}$ (surface activity of a glassy carbon electrode surface). Finally, we discuss the implications of our findings for voltammetric
    experiments more generally.
\end{abstract}

\section{Introduction}

In a previous paper \cite{gavaghan2017}, we described the use of Bayesian
inference for quantitative comparison of voltammetric methods for investigating
electrode kinetics. We illustrated the utility of the approach by comparing the
information content in both dc and ac voltammetry at a planar electrode for the
case of a quasi-reversible one electron reaction mechanism. Using both synthetic
and experimental data, we were able to demonstrate that realistic levels of
purely random experimental (Gaussian) noise have a relatively minor affect on
the inverse problem of recovering  both the faradaic (the reversible formal
potential, $E_0$, the standard heterogeneous charge transfer rate constant
$k_0$, and the charge transfer coefficient $\alpha$) and non-faradaic
(uncompensated resistance, $R_u$, and double layer capacitance, $C_{dl}$)
parameters that govern this reaction mechanism. We also demonstrated the clear
advantages in terms of accuracy of parameter recovery of the large amplitude ac
approach. With this ability of being able to recover parameter values reliably
from a single experimental data set in place, we are now in a position to go on
to investigate \emph{system level} variability i.e. if we repeat the same
experiment multiple times and implement our Bayesian parameter recovery
procedure for each data set independently, how reproducibly do we recover the
governing parameters?

To approach this problem, we again return to the ``pathological"
$[$Fe(CN)$_6]^{3-/ 4-}$ process. It is well known that for this system the
kinetic parameters reported are highly variable even when using apparently
identical electrodes and conditions (see
\cite{bond2005changing,A_mccreery2008advanced,B_mccreery2012comment,C_patel2012new,D_ji2006oxygenated,E_curulli2005kinetic,F_mundinamani2014cyclic,G_dekanski2001glassy,H1_gonccalves2006electrochemical,H2_wightman1984methods,I_patel2012new,J_kuwana2018analytical,K_kuwana2018analytical}
and references cited
therein).
Using our new approach we are able to show that this difficulty is not due to the impact of
random experimental noise, since for each individual data set we are able to fit
the mathematical model to the experimental data extremely accurately. However,
we are able to demonstrate that the recovered values  from each individual data
set vary significantly from one another, i.e. the system itself varies between
experimental runs, and so the recovered parameter values are extremely sensitive
to the precise experimental conditions pertaining on that particular run. Since
we have already demonstrated its advantages over the dc approach, we restrict
ourselves to the ac case in this paper.

\section{Methods}

\subsection{Experimental methods \label{exptl}}

Details of the experimental data sets used in this paper have been given
previously in \cite{bond2005changing,Morrisetal}. In summary, large amplitude ac
voltammetry was performed in a standard three-electrode cell, using a glassy
carbon macrodisk (diameter 3 mm) working electrode. All potentials are reported versus an
Ag/AgCl/KCl(3 M) reference electrode (hereinafter Ag/AgCl). The frequencies, amplitude and scan rate were 9.02 Hz, 0.080 V and 0.894 $\text{Vs}^{-1}$ respectively and data were
collected over the potential range of 0.5 to -0.1 V versus Ag/AgCl.  The surface
area of the electrode was estimated as $0.070\;\mathrm{cm}^2$. The value of the diffusion coefficient, $D$, of
$\left[\mathrm{Fe(CN)}_6\right]^{3-}$ was found to be $7.2
\times 10^{-6}\;\mathrm{cm}^2\; \mathrm{s}^{-1}$, as described in
\cite{Morrisetal}. In this paper we make use of the ten repeated data sets for
ac voltammetry taken from \cite{Morrisetal} for the reduction of aqueous 1.0 mM
$[$Fe(CN)$_6]^{3-}$ in 3 M KCl aqueous electrolyte solution. In order to reduce
the amount of computation required for the parameter inference, we use a moving
average window of length 21 to reduce the number of experimental data points
within each data set to about 25,000 data points.

\subsection{Mathematical Modelling}

The details of the mathematical and computational modelling approach that we
have taken in this paper were given previously in \cite{Morrisetal,gavaghan2017}. In summary, our chosen experimental system is modelled as a quasi-reversible reaction

\begin{align} \label{reaction}
A + e^- \cee{&<=>[E_0,k_0,\alpha]} B,
\end{align}

\noindent where species $A$ and $B$ are in solution, and $E_0$,
$k_0$, and $\alpha$  are the reversible formal potential, standard heterogeneous
charge
transfer rate constant at $E_0$ and the charge transfer coefficient, respectively. We assume
that the Butler-Volmer formalism applies to the electron transfer
process \cite{BardFaulkner,brett1993principles,pletcher2001instrumental}. We also
assume that both convection and migration can be neglected since we are using a
macrodisk stationary electrode and an excess of supporting electrolyte,
respectively. Since we assume equal diffusion coefficients for each species $A$
and $B$ ($D_A = D_B = D$) we need to solve for the concentration of only one of the
species  (i.e. the concentrations $c_A$, $c_B$ of species $A$ and $B$ satisfy
$c_A   = c_{\infty} - c_B $, where $c_\infty$ is the bulk concentration of
species $A$) and we choose to solve for species $A$. We can then use Fick's
second law to model the variation with time of species $A$ via

\begin{align}
\frac{\partial c_A}{\partial t} &= D \frac{\partial^2 c_A}{\partial x^2}, \label{diffeqtn}
\end{align}

\noindent where $x$ is distance from the electrode surface and $t$ is time. The
initial and boundary conditions are

\begin{align}
c_A(x,0) &= c_{\infty} \nonumber \\
c_A &\rightarrow c_{\infty},  \quad \text{as} \quad x \rightarrow \infty,\quad t>0. \label{icsandbcs}
\end{align}

At the electrode surface, $x=0$, for $t>0$, we have the conservation and flux conditions

\begin{align}
D \frac{\partial c_A}{\partial x}=\frac{I_f}{FS}, \label{fluxatelectrode}
\end{align}

\noindent along with the Butler-Volmer condition

\begin{align}
    D \frac{\partial c_A}{\partial x} = \text{ }&k_0
    \left[(c_{\infty}-c_A)\exp\left( (1-\alpha) \frac{F}{RT}(E_{\mbox{\tiny
    eff}}(t)-E_0)\right)\right.
    \left.-c_A\exp\left(-\alpha\frac{F}{RT}(E_{\mbox{\tiny
    eff}}(t)-E_0)\right)\right]. \label{BVeqtn}
\end{align}

Here, $I_f$ is the faradaic current, $S$ is the electrode area, and
$E_{\mbox{\tiny eff}}(t)$ is the {\em effective} applied potential (defined
below).

We complete the model by defining $E_{\mbox{\tiny{app}}}(t)$ to be the applied
potential, then for the case of an ac voltammetry ramp we have

\begin{align}
E_{\mbox{\tiny{app}}}(t) =  E_{\mbox{\tiny{start}}}  \left\lbrace
      \begin{array}{ll}
       + v t + \Delta E \sin {(\omega t)}, \qquad 0 \le t \le t_{\mbox{\tiny
        reverse}}, \\
        -  v t + 2 v t_{\mbox{\tiny reverse}} + \Delta E \sin {(\omega t)},
        \qquad  t_{\mbox{\tiny reverse}}\le t \le 2t_{\mbox{\tiny reverse}}
      \end{array}
  \right.
 \label{Eac}
\end{align}

\noindent where $v$ is the sweep rate,  $E_{\mbox{\tiny{start}}}$ is the initial
potential, $t_{\mbox{\tiny reverse}}$ is the time of switching from the forward to the reverse sweep in cyclic voltammetry, $\omega$ is the radial frequency and $\Delta E$ is the amplitude of the sine wave. The {\em effective} applied potential can now be defined as

\begin{align}
E_{\mbox{\tiny{eff}}}(t) = E_{\mbox{\tiny{app}}}(t) - E_{\mbox{\tiny{drop}}} = E_{\mbox{\tiny{app}}}(t) - I_{\mbox{\tiny{tot}}} R_u \nonumber
\end{align}

\noindent where $E_{\mbox{\tiny{drop}}}$ models the effect of uncompensated
resistance,
$R_u$. $I_{\mbox{\tiny{tot}}}$ is the total (measured) current, and combines the
faradaic current and the background capacitive current, $I_c$, which can be
modelled as

\begin{align}
I_c &= C_{dl}\frac{dE_{\mbox{\tiny{eff}}}}{dt}, \label{Ic}
\end{align}
where $C_{dl}$ is the double layer capacitance (assumed constant in this work), and then
\begin{align}
I_{\mbox{\tiny{tot}}} = I_f + I_c. \label{Itot}
\end{align}

Equations \ref{diffeqtn}--\ref{Itot} are non-dimensionalised as described
previously \cite{Morrisetal, gavaghan2017}. The resulting non-dimensional system
of equations is solved using an implicit finite difference method with an
exponentially expanding grid (again, as described previously \cite{Sheretal}).
We can now see mathematically that the reaction mechanism in Eq. \ref{reaction}
is governed by five parameters $(E_0, k_0, \alpha, C_{dl}, R_u)$, and we will
collectively denote these parameters by the vector $ \boldsymbol\theta$. The
{\em inverse} problem that we wish to solve can be defined as finding the best
possible approximation to $\boldsymbol\theta$ given our measured experimental
output trace of the current $I_{\mbox{\small{\tiny{tot}}}}^{\mbox{\tiny{data}}}$
versus potential.

\section{Parameter Recovery Methods}

In a recent paper we described in detail how methods of Bayesian inference can
be used to solve the inverse problem of parameter recovery \cite{gavaghan2017}
from voltammetric data. We illustrated how these methods yield not only a point
estimate for each parameter of interest, but also a measure of our confidence in
that estimate. Full details of the approach that we adopted, including the
algorithms used can be found in \cite{gavaghan2017}, so that here we simply give
a brief outline of the methods.

\subsection{Bayesian Inference}

We denote by $\mathbf{y} = (y_1,\dots,y_T)$ the experimental data trace, that
is, the total measured current $I_{\mbox{\tiny{tot, t}}}^{\mbox{\tiny{data}}}$
at each time point $t$. In our previous work (see Figure 4 and Table S2 of
\cite{Morrisetal}) we demonstrated that, to a very good approximation,
experimental measurements can be assumed to be subject to normally distributed
random noise, typically with zero mean and some standard deviation which we
will denote by $\sigma$ (we showed that typical values of the standard deviation
of the experimental noise are in the range 1 to 2\% of the peak current). Our
mathematical model of the system, Eqs \ref{diffeqtn} to \ref{Itot} above,
assumes that the observed experimental data $\mathbf{y}$ is a function of the
parameters of interest  $\boldsymbol{\theta}$,

$$
\boldsymbol{\theta} = (E_0,k_0,\alpha,C_{dl},R_u).
$$

We now assume further that the parameters $\boldsymbol{\theta}$ are themselves
drawn from a probability distribution. We can then frame our inverse problem as
trying to find this probability distribution for  $\boldsymbol{\theta}$ {\em given the observed
values of the data} $\mathbf{y}$, and denote this probability distribution as $P(\boldsymbol{\theta} \vert \mathbf{y})$ (the vertical line indicates that the values of  $\mathbf{y}$ are {\em given}). In Bayesian inference, $P(\boldsymbol{\theta}\vert\mathbf{y})$ is termed the {\em posterior probability density} or {\em posterior distribution}; this is the distribution that we want
to approximate.

We now make use of Bayes' rule which states
\begin{equation}
P(\boldsymbol{\theta} \vert \mathbf{y}) = \frac{P(\mathbf{y}\vert\boldsymbol{\theta}) P(\boldsymbol{\theta})}{P(\mathbf{y})},
\label{eq:bayes}
\end{equation}

\noindent where $P(\boldsymbol{\theta})$ is called the prior distribution of
$\boldsymbol{\theta}$ and is chosen to capture any prior knowledge we have of
$\boldsymbol{\theta}$ before any experimental observation.  The distribution
$P(\mathbf{y}\vert\boldsymbol{\theta})$ is the probability density of the
experimental data $\mathbf{y}$ given a model parameterised with parameters
$\boldsymbol{\theta}$, and is termed the {\em likelihood} of the data; assuming
a known distribution of the error in the data this likelihood can be calculated.
$P(\mathbf{y})$ is a normalising term (which is the integral of all possible
densities $P(\mathbf{y},\boldsymbol{\theta}) = P(\mathbf{y} \vert
\boldsymbol{\theta})P(\boldsymbol{\theta})$ over all values of $\boldsymbol{\theta}$), and ensures that the posterior density $P(\boldsymbol{\theta}\vert\mathbf{y})$ integrates to $1$. In practice, the calculation of this normalising term (which can be very computationally expensive) is avoided by considering ratios of the likelihood (see below).

\subsection{Calculating the likelihood \label{section:likelihood}}
Writing the likelihood as
\begin{equation}
L(\boldsymbol{\theta} \vert \mathbf{y}) = P(\mathbf{y}\vert\boldsymbol{\theta}),
\end{equation}
we can re-arrange Bayes' rule in Eq.~\ref{eq:bayes} to give
\begin{equation}
P(\boldsymbol{\theta} \vert \mathbf{y}) \propto P(\boldsymbol{\theta})L(\boldsymbol{\theta}\vert\mathbf{y}).
\label{eqn:posterior_proportional_to_prior_and_likelihood}
\end{equation}
Since we assume that the errors are independent at each time point, the conditional probability density of observing the experimental trace from $t=1,\ldots,T$ given $\boldsymbol{\theta}$ is simply the product of the probability density functions at each time point, that is, the likelihood is given by

\begin{align}
  L(\boldsymbol{\theta} \vert \mathbf{y}) &= \prod_{t=1}^{T}P(y_t \vert
  \boldsymbol{\theta})
= \prod_{t=1}^{T} \mathcal{N}(y_t \vert f_t(\boldsymbol{\theta}),\sigma^2) \\ &=
  \prod_{t=0}^{T} \frac{1}{\sqrt{2 \pi \sigma^2}} \exp
  \left(-\frac{\left(y_t-f_t(\boldsymbol{\theta})\right)^2}{2\sigma^2}\right),
\label{eqn:product_likelihood}
\end{align}

\noindent using the assumption that the experimental noise is normally
distributed with a
mean zero and variance of $\sigma^2$. For notational simplicity we have set
$f_t(\boldsymbol{\theta}) = I_{\mbox{\tiny{tot}},t}^{\mbox{\tiny{model}}}$.

Since, in most experimental situations in electrochemistry, {\em  a priori} we
will have only a rough idea of what the values of the parameters are likely to
be, we assume an ``uninformative" prior and use a uniform distribution for each
parameter across a suitably wide range, that is

\begin{equation} \label{eq:prior}
P(\boldsymbol\theta) = \left\{
\begin{array}{l}
c,\:\{\boldsymbol\theta\} \text{\:in\:some\:suitably\:chosen\:5-dimensional\: hypercube,}\\ 0,\text{\:otherwise},
\end{array}
\right.
\end{equation}

\noindent where $c$ is a non-zero finite normalizing constant. The bounds for
this hypercube were set to

\begin{align*}
    E_{reverse} + 0.1\delta E &\le E_0 \le
    E_{start}-0.1\delta E,\\
    0 &\le k_0 \le 1 \text{ cm s}^{-1},\\
    0.4 &\le \alpha \le 0.6,\\
    0 &\le C_{dl} \le 200\text{ }\mu\text{F cm}^{-2},\\ 0 &\le R_u \le 80
    \text{ }\Omega,
\end{align*}

\noindent where $\delta
    E=E_{start}-E_{reverse}$.

Note that this prior is only used for the single level Markov Chain Monte Carlo (MCMC) algorithm; in the hierarchical MCMC algorithm, this is replaced by a multivariate normal, as described below. However, these bounds are still used in the hierarchical MCMC algorithm to prevent the lower level samplers from accepting any samples that lie outside these bounds.

\subsubsection{Markov Chain Monte Carlo parameter inference}\label{sec:MCMC}

To obtain a sample from the posterior distribution $P(\boldsymbol{\theta} \vert
\mathbf{y})$ we make use of the Markov Chain Monte Carlo method. In
outline, this involves finding an approximation to the posterior distribution
$P(\boldsymbol{\theta} \vert \mathbf{y})$ by drawing a finite (but sufficiently
large to be accurate) number of samples from this distribution. To do this we
simulate a
Markov Chain whose limiting distribution is the required posterior distribution
using an efficient implementation of the \textit{Metropolis-Hastings} algorithm
\cite{MetropolisHastingsref}, within which candidate parameter sets are proposed
from a \textit{proposal distribution} $q(\boldsymbol{\theta_{cand}} \vert
\boldsymbol{\theta_i})$ which depends only on the previously accepted parameter
set $\boldsymbol{\theta_i}$; we take $q(\boldsymbol{\theta_{cand}} \vert
\boldsymbol{\theta_i})$ to be a multivariate normal distribution. If the
proposed parameter set contains any parameters outside the range of the prior,
then the parameter set is assigned an acceptance probability of $0$, i.e. it is
rejected, and the previously accepted parameter set is added to the Markov chain
--- that is, $\boldsymbol{\theta_{i+1}}=\boldsymbol{\theta_i}$. Otherwise, we
compare $\boldsymbol{\theta_{cand}}$ to the current parameter set
$\boldsymbol{\theta_i}$ by calculating the ratio of the posteriors of the two
parameter sets. If the candidate parameter set has a greater posterior density
value than the existing parameter set then it will be added to the Markov chain,
that is $\boldsymbol{\theta_{i+1}} = \boldsymbol{\theta_{cand}}$. Otherwise,
(making use of Eq.\  \ref{eqn:posterior_proportional_to_prior_and_likelihood})
the proposed parameter is accepted with probability, $r$, given by

\begin{equation}\label{eq:acceptance_prob}
  r = \min \Bigg\{ \frac{P(\boldsymbol{\theta_{cand}})
  L(\boldsymbol{\theta_{cand}} \vert \mathbf{y})}{P(\boldsymbol{\theta_i})
  L(\boldsymbol{\theta_i} \vert \mathbf{y})},1 \Bigg\}.
\end{equation}

If the
proposed parameter set is rejected (with probability $1-r$), then the previously
accepted parameter set is again added to the Markov chain --- that is,
$\boldsymbol{\theta_{i+1}}=\boldsymbol{\theta_i}$.

Since the likelihoods for large samples are extremely small, in practice we work
with the natural log of the likelihood which reduces to

\begin{equation}
 l(\boldsymbol{\theta} \vert \mathbf{y}) = - T\log(\sigma) - \frac{1}{2\sigma^2}\sum_{t=1}^T (y_t - f_t (\boldsymbol{\theta}))^2, \label{loglikelihood}
\end{equation}

\noindent where terms which are constant in $\boldsymbol\theta$ have been
removed (since
these will cancel on taking the difference of log-likelihoods in the
Metropolis-Hastings algorithm). More comprehensive descriptions of the theory of
MCMC can be found in the statistics literature (see for example
\cite{gilks1996}).

\subsection{Practical implementation of the Metropolis-Hastings Algorithm}

In practice, we make use of an adaptive covariance matrix version of the
Metropolis-Hastings algorithm which helps identify the directions in parameter
space which have the highest likelihood values \cite{haario2001}. At each
iteration of the algorithm, the covariance matrix of the multivariate normal
distribution is updated and a scalar value is also updated to define the width
of the distribution. In the results presented in this paper, we run our MCMC
chains for 10,000 samples and discard the first 5,000 samples as `burn in' (see
\cite{gilks1996}). To ensure efficiency in our Monte-Carlo sampling, we first
find the location of the optimum, that is, the maximum likelihood estimate of
$\boldsymbol\theta$  using a standard global minimisation algorithm (we use the
{\em cma-es} algorithm \cite{auger2005restart}), which we use as a seed point
for the MCMC algorithm as described in \cite{gavaghan2017}. In the results
section, these samples are shown as histograms which illustrate the nature of
the posterior distribution.

\subsection{Hierarchical Bayesian Inference}

In the results section, the MCMC algorithm described above is used to estimate
the parameter values and their posterior distributions from each of the ten
repeat runs of the ac voltammetry experiment for the reduction of aqueous 1 mM
$[$Fe(CN)$_6]^{3-}$, as described in the Experimental methods section.  This
will allow us to
show the degree of variability in the recovered values of the parameters across
these ten data sets. This will in turn allow us to postulate that on each repeat
of the experiment the parameters  $\boldsymbol\theta = \left\{E_0, k_0, \alpha,
C_{dl}, R_u \right\}$ are themselves drawn from a multivariate normal
distribution with mean hyper-parameters $\boldsymbol\mu = (\hat{E_0}, \hat{k_0},
\hat{\alpha}, \hat{C_{dl}},\hat{R_u})$ and a covariance hyper-parameter matrix
$\boldsymbol\Sigma$.  Our aim now is to sample from the distributions of
$\boldsymbol\mu$ and
$\boldsymbol\Sigma$ to enable us to quantify how $\boldsymbol\theta$ varies
between different experiments.

We first write the posterior distribution of all of our parameters given the
data, taking into account our new hierarchical model structure

\begin{align}
P(\boldsymbol\mu,\boldsymbol\Sigma,\boldsymbol\theta_1,...,\boldsymbol\theta_n, \vert \mathbf{y}_1,...,\mathbf{y}_n)
  &\propto P(\boldsymbol\mu,\boldsymbol\Sigma,\boldsymbol\theta_1,...,\boldsymbol\theta_n)
  \prod_{i=0}^n P(\mathbf{y}_i \vert\boldsymbol{\theta}_i), \\
  &= P(\boldsymbol\mu,\boldsymbol\Sigma)
  \prod_{i=0}^n P(\boldsymbol{\theta}_i \vert \boldsymbol\mu,\boldsymbol\Sigma)
  \prod_{i=0}^n P(\mathbf{y}_i \vert\boldsymbol{\theta}_i),
\end{align}

\noindent where $\boldsymbol\theta_1,...,\boldsymbol\theta_n$ are all the bottom level
parameters for the $n$ different experiments, and
$\mathbf{y}_1,...,\mathbf{y}_n$ are the corresponding measurements.

We choose a multivariate normal distribution for the bottom level
parameters $\boldsymbol\theta$

\begin{equation}\label{eq:normal}
  P(\boldsymbol\theta_i| \boldsymbol\mu,\boldsymbol\Sigma) =
  \mathcal{N}(\boldsymbol\mu,\boldsymbol\Sigma).
\end{equation}

It now remains to choose a suitable hyper-prior
$P(\boldsymbol\mu,\boldsymbol\Sigma)$, which for ease of computation is
generally taken to be a normal-inverse-Wishart distribution (see for example
\cite{huang2010hierarchical})

\begin{equation}\label{eq:niw}
  P(\boldsymbol\mu,\boldsymbol\Sigma) =
  \mathcal{NIW}(\boldsymbol{\mu}_0,\kappa_0,\nu_0,\boldsymbol\Psi).
\end{equation}

The distributions in Eqs.~\ref{eq:normal} and \ref{eq:niw} are chosen to be
conjugate, so that the conditional distribution of the hyper-parameters can be
analytically derived as
\cite{murphy2007conjugate}

\begin{equation}\label{eq:niw_post}
  P(\boldsymbol\mu,\boldsymbol\Sigma|\boldsymbol\theta_1,...,\boldsymbol\theta_n) =
  \mathcal{NIW}(\frac{\kappa_0\boldsymbol\mu_0 + n\hat{\boldsymbol{\theta}}}{\kappa_0 + n},
               \kappa_0 + n,\nu_0 + n,\boldsymbol\Psi + \mathbf{C} +
               \frac{\kappa_0n}{\kappa_0+n}(\hat{\boldsymbol{\theta}}-\boldsymbol\mu_0)(\hat{\boldsymbol{\theta}}-\boldsymbol\mu_0)^T),
\end{equation}

\noindent where $\hat{\boldsymbol{\theta}}$ and $\mathbf{C}$ are the sample mean
and covariance of the bottom level parameters

\begin{align}\label{eq:niw_post2}
\hat{\boldsymbol{\theta}}&=\frac{1}{n} \sum_{i=0}^n \boldsymbol\theta_i, \\
\mathbf{C} &=
\sum_{i=0}^n (\boldsymbol\theta_i-\hat{\boldsymbol\theta})
(\boldsymbol\theta_i-\hat{\boldsymbol\theta})^T.
\end{align}

We can now use our original adaptive MCMC algorithm to sample from the bottom
level parameters, combined with classical Gibbs sampling and
Eq.~\ref{eq:niw_post} to sample from the hyper-parameters (see Algorithm
\ref{alg:metropolis_within_gibbs} for details). For the hyperprior parameters we
use $\kappa_0=0$ and $\nu_0=1$. We set $\boldsymbol\mu_0$ to be the centre
point of the 5-dimensional hypercube in Eq.~\ref{eq:prior}, and
$\boldsymbol\Psi$ as a diagonal matrix with the standard deviation of each
parameter set to half the width of this same hypercube. To prevent the bottom
level MCMC chains from exploring unphysical parameter regimes, we automatically
reject any proposed point which lies outside of the hypercube.

 \begin{algorithm}
  \caption{Metropolis within Gibbs}
  \label{alg:metropolis_within_gibbs}
  \begin{algorithmic}
  \STATE $s=0$
  \STATE $\boldsymbol\mu_s,\boldsymbol\Sigma_s = \mbox{SampleFrom}\mathcal{NIW}(\boldsymbol{\mu}_0,\kappa_0,\nu_0,\boldsymbol\Psi)$ \COMMENT{using Eq.~\ref{eq:niw}}
  \REPEAT
    \FOR{$i=1$ \TO $n$}
      \STATE $\boldsymbol\theta_i =
      \mbox{AdaptiveMCMCStep}(\boldsymbol\mu_s,\boldsymbol\Sigma_s)$
      \COMMENT{see Eq.~\ref{eq:acceptance_prob} and enclosing section}
    \ENDFOR
    \STATE $s = s + 1$
    \STATE $\boldsymbol\mu_s,\boldsymbol\Sigma_s = \mbox{SampleFrom}\mathcal{NIW}(\boldsymbol{\mu}_0,\kappa_0,\nu_0,\boldsymbol\Psi,\boldsymbol\theta_1,...,\boldsymbol\theta_n)$ \COMMENT{using Eq.~\ref{eq:niw_post} \& \ref{eq:niw_post2}}
  \UNTIL{finished sampling}
  \end{algorithmic}
  \end{algorithm}

\subsection{Generating synthetic data as a test case}

To test our inference procedure and algorithms
we make use of ``synthetic" test data i.e. we generate ten synthetic data sets
by
solving Equations~\ref{diffeqtn} to \ref{Itot} for a randomly chosen set of
values of $\boldsymbol{\theta} = (E_0, k_0, \alpha, C_{dl}, R_u)$, drawn from a
multivariate normal distribution with mean $\boldsymbol{\theta}_{\mbox{\tiny
true}} = (7.27,2.01, 0.53,3.70\times 10^{-3}, 1.06\times 10^{-2})$ (in
non-dimensional units), and standard deviation $\boldsymbol{\sigma}_{\mbox{\tiny
true}} = ( 0.06,0.7, 0.005,0.7\times 10^{-3}, 0.3\times 10^{-2})$. We then add
randomly
generated Gaussian noise at each time point with zero mean and standard
deviation of 0.3\% of the maximum current (chosen to match the
experimental noise level).

\section{Results}

\subsection{Synthetic data}

\begin{figure*}[htbp]
\begin{centering}
\includegraphics[width= \textwidth]{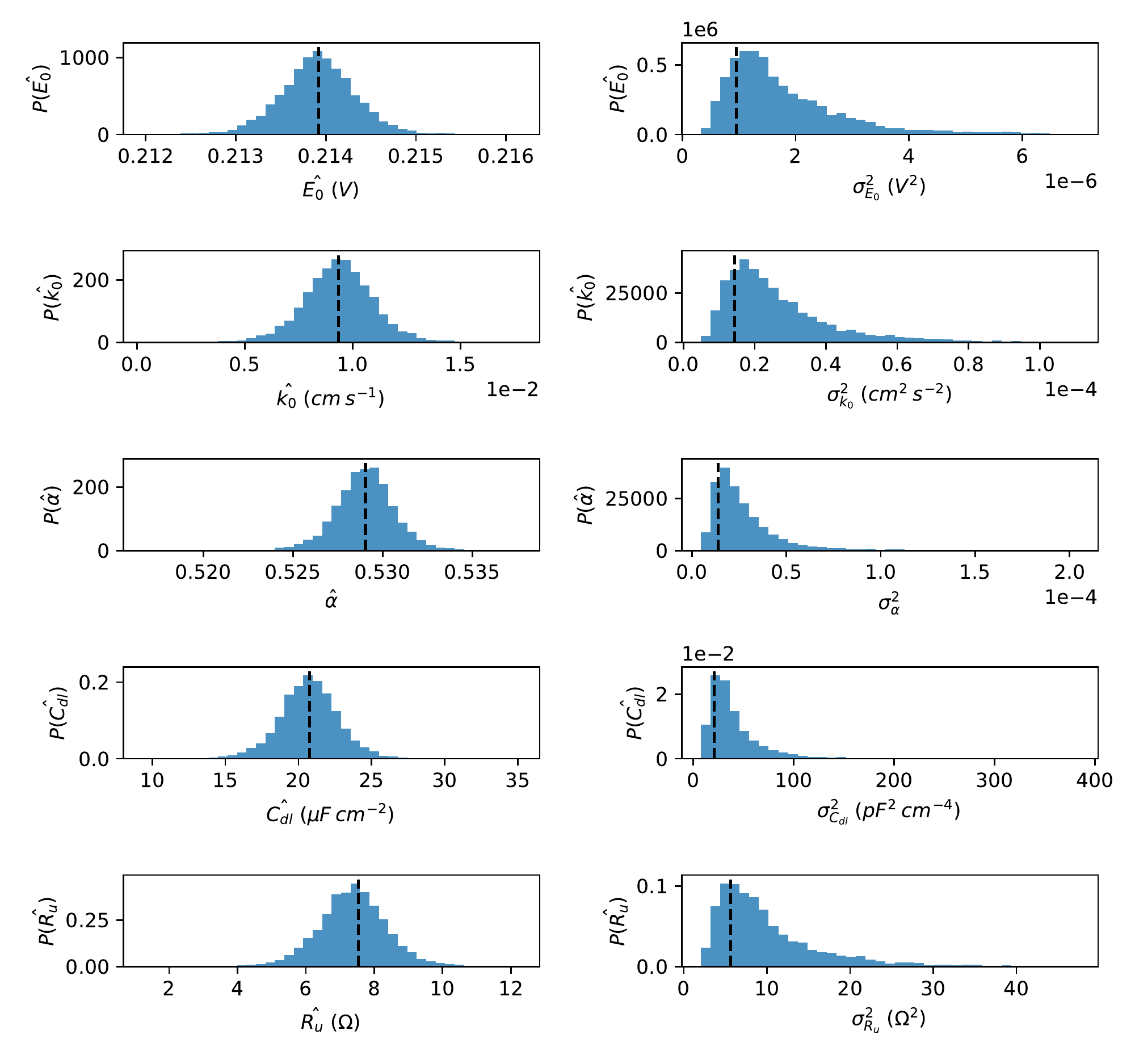}
\end{centering}
  \caption{Histograms of sampled hyper-parameter distributions $\boldsymbol\mu$
  (left column) and the variances of $\boldsymbol\Sigma$ (right column),
  generated by using
  Algorithm \ref{alg:metropolis_within_gibbs} on 10 sets of synthesised ac
  voltammetric data. The dashed lines indicate the sample mean and sample
  standard deviation of the true parameters used to generate the synthetic
  datasets, and since this aligns correctly with the maximum likelihood point of
  the histograms we can be confident that our implementation of Algorithm
  \ref{alg:metropolis_within_gibbs} is sampling the correct posterior
  distribution for $\boldsymbol\mu$ and $\boldsymbol\Sigma$.}
  \label{fig:synthetic_histograms}
\end{figure*}

To verify that we have implemented our hierarchical Bayes algorithms correctly
we briefly describe the results of using synthetic data to test our inference
procedure. Figure \ref{fig:synthetic_histograms} shows the sampled distributions
of the hyper-parameters. The dashed lines indicate the expected peak of each
distribution, as calculated from the sample mean and variance of the
ten true values of $\boldsymbol{\theta}_{\mbox{\tiny true}}$. As can be seen the
hierarchical MCMC algorithm obtains the correct peak for all the five different
mean and variance hyper-parameters.

\subsection{Experimental data}

\begin{figure*}[htbp]
\begin{centering}
\includegraphics[width= \textwidth]{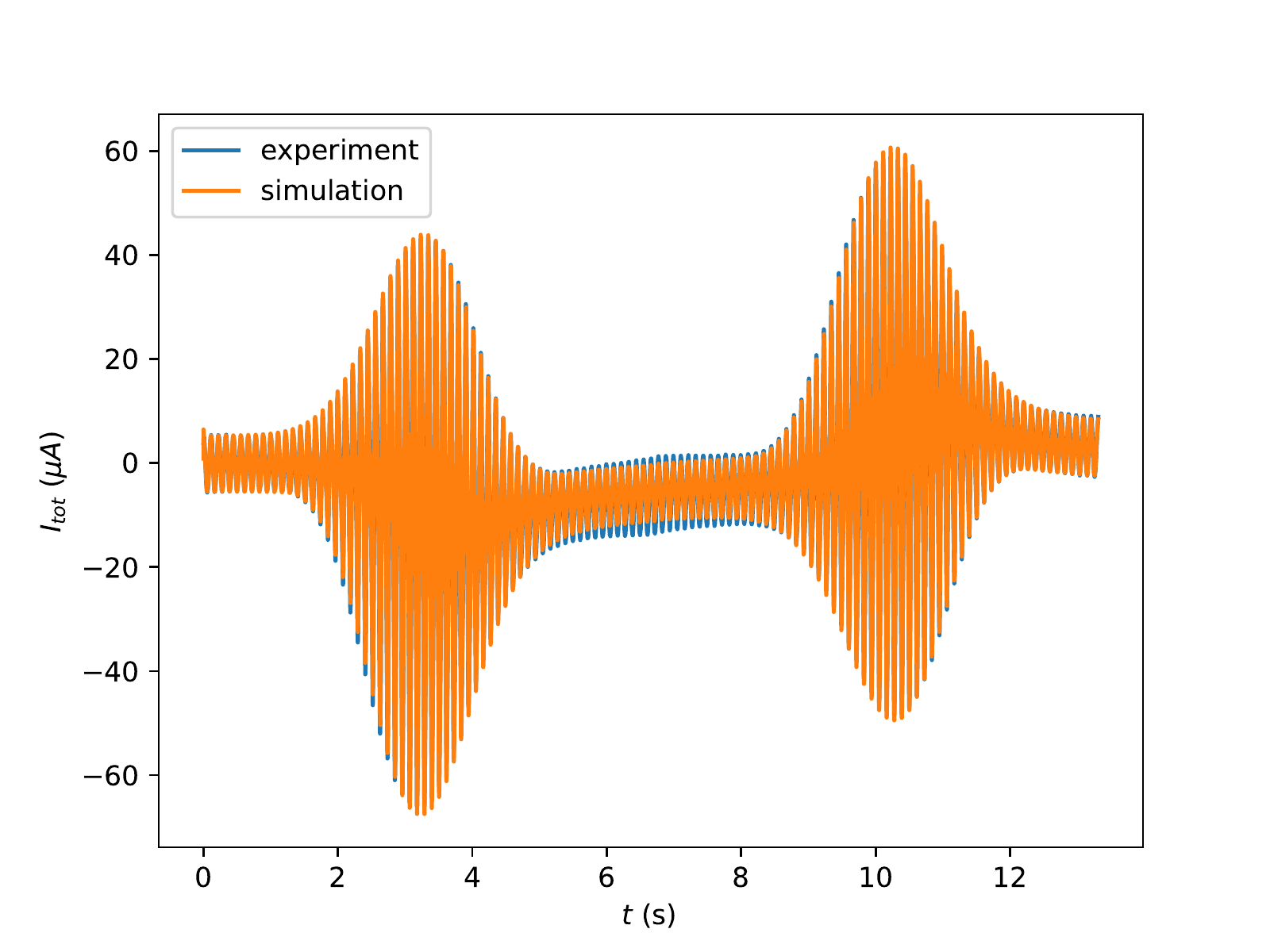}
\end{centering}
  \caption{Comparison between the simulated current trace and the experimental data for the reduction of aqueous 1 mM
  $[$Fe(CN)$_6]^{3-}$ (dataset 1 from reference \cite{Morrisetal}).  Simulation
  parameters were obtained by maximum likelihood estimate, and their values in
  dimensional units were $E_0=0.214$ V, $k_0=0.010$
     $\text{ cm s}^{-1}$, $\alpha=0.528$, $C_{dl}=16.9$ $\mu \text{F }
     \text{cm}^{-2}$, $R_u=0.00$ $\Omega$. }
  \label{fig:current_traces}
\end{figure*}

Figure \ref{fig:current_traces} shows an example of the initial fitting process
for experimental data set 1.  Once the best fit for each data set was obtained
using maximum likelihood estimation and the \textit{cma-es} optimisation
algorithm, each
of the bottom level MCMC algorithms was initialised at these points. Then
Algorithm \ref{alg:metropolis_within_gibbs} was used to generate 5,000 samples
(10,000 total samples, of which the first 5,000 was discarded as burn-in) of the
parameters $\boldsymbol\theta_i$ and hyper-parameters $\boldsymbol\mu$ and
$\boldsymbol\Sigma$.

\begin{figure*}[htbp]
\centering
\includegraphics[width= 0.49\textwidth]{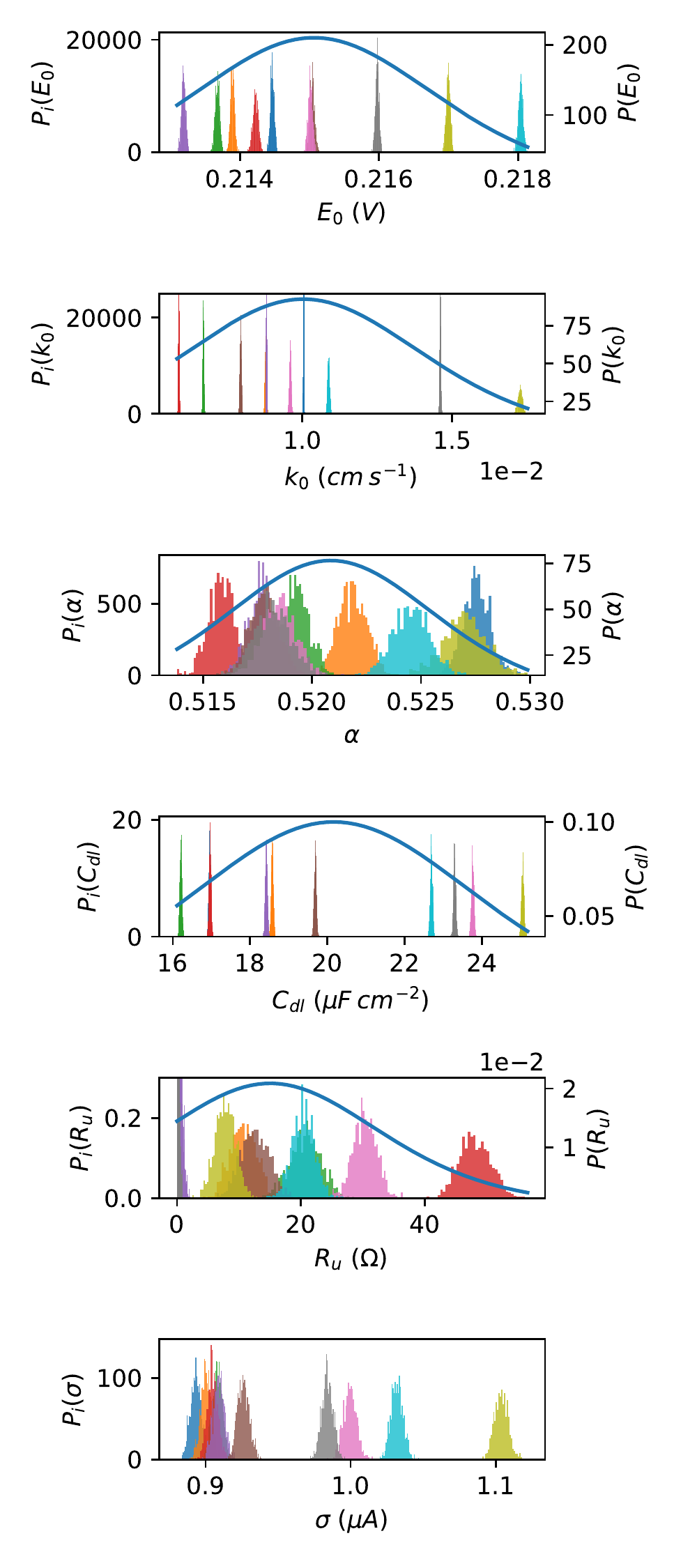}
\includegraphics[width= 0.49\textwidth]{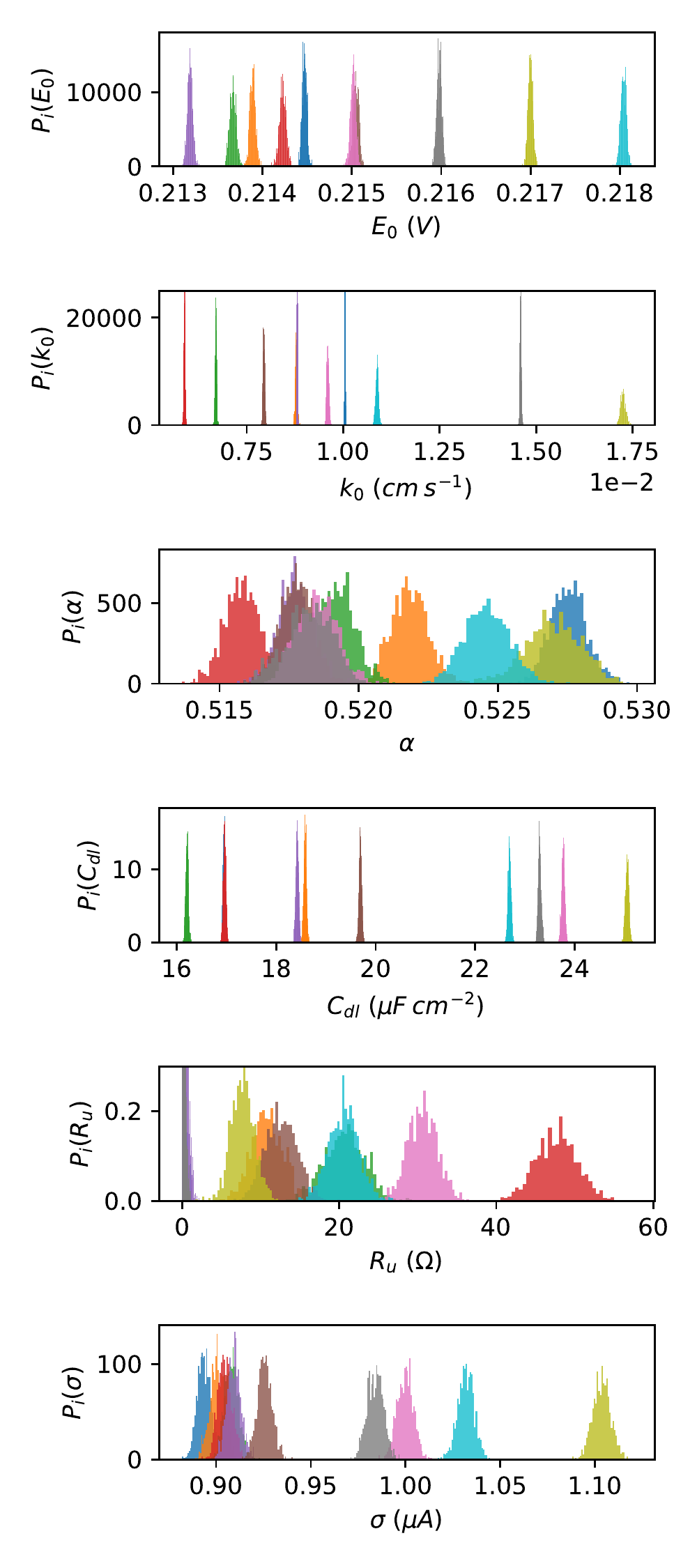}
  \caption{Analysis of ten independent ac voltammetric experiments for the
  reduction of aqueous 1 mM $[$Fe(CN)$_6]^{3-}$ using the hierarchical MCMC
  algorithm \ref{alg:metropolis_within_gibbs}. The samples obtained from the 10
  lower level adaptive MCMC samplers (i.e. $\boldsymbol\theta_i$) are shown as
  histograms with the axis label $P_i(\cdot)$, and the different chains from
  $i=1,..,10$ are shown with different
  colours. The left column plots show the samples obtained using the
  hierarchical model, and these also show the posterior predictive
    distribution $P(\cdot)$ for each variable, calculated by summing the
    Gaussian distributions described by the samples of the
    hyper-parameters $\boldsymbol\mu$ and  $\boldsymbol\Sigma$ (see Figure
    \ref{fig:htrace} for histograms of these hyper-parameter
    samples).  For comparison, the right column plots show the
    samples of $\boldsymbol\theta_i$ with no hierarchical model (and thus no
    hyper-parameter samples). This is identical to the analysis done in our
    previous paper \cite{gavaghan2017}. }
  \label{fig:bottom_level_chains}
\end{figure*}

Figure \ref{fig:bottom_level_chains} (left column) shows histograms of the
samples of $\boldsymbol\theta_i$ taken from the bottom level samplers in
Algorithm \ref{alg:metropolis_within_gibbs}. On the
same axis is drawn the posterior predictive distribution for each variable,
calculated by summing the individual Gaussian distributions described by each
sample of the hyper-parameters $\boldsymbol\mu$ and  $\boldsymbol\Sigma$.
The posterior predictive distributions describes the distribution of
each parameter that would be expected from another repeat of the experiment,
given the results of the 10 already observed experiments. As can be seen, this
distribution covers the width of all 10 bottom level samples of
$\boldsymbol\theta_i$, and illustrates the significantly greater variation in
each parameter that is expected between subsequent experiments.

It is interesting to compare the samples obtained from the hierarchical model to
those taken using the original non-hierarchical model (i.e. just running ten
independent MCMC chains on the ten different data sets), and this is shown in
Figure \ref{fig:bottom_level_chains} (right column). As can be seen, the
result in this case are almost identical to the hierarchical model, giving us
confidence that we are capturing the distributions of $\boldsymbol\theta_i$
correctly in each case.

\begin{figure*}[htbp]
\centering
\includegraphics[width= \textwidth]{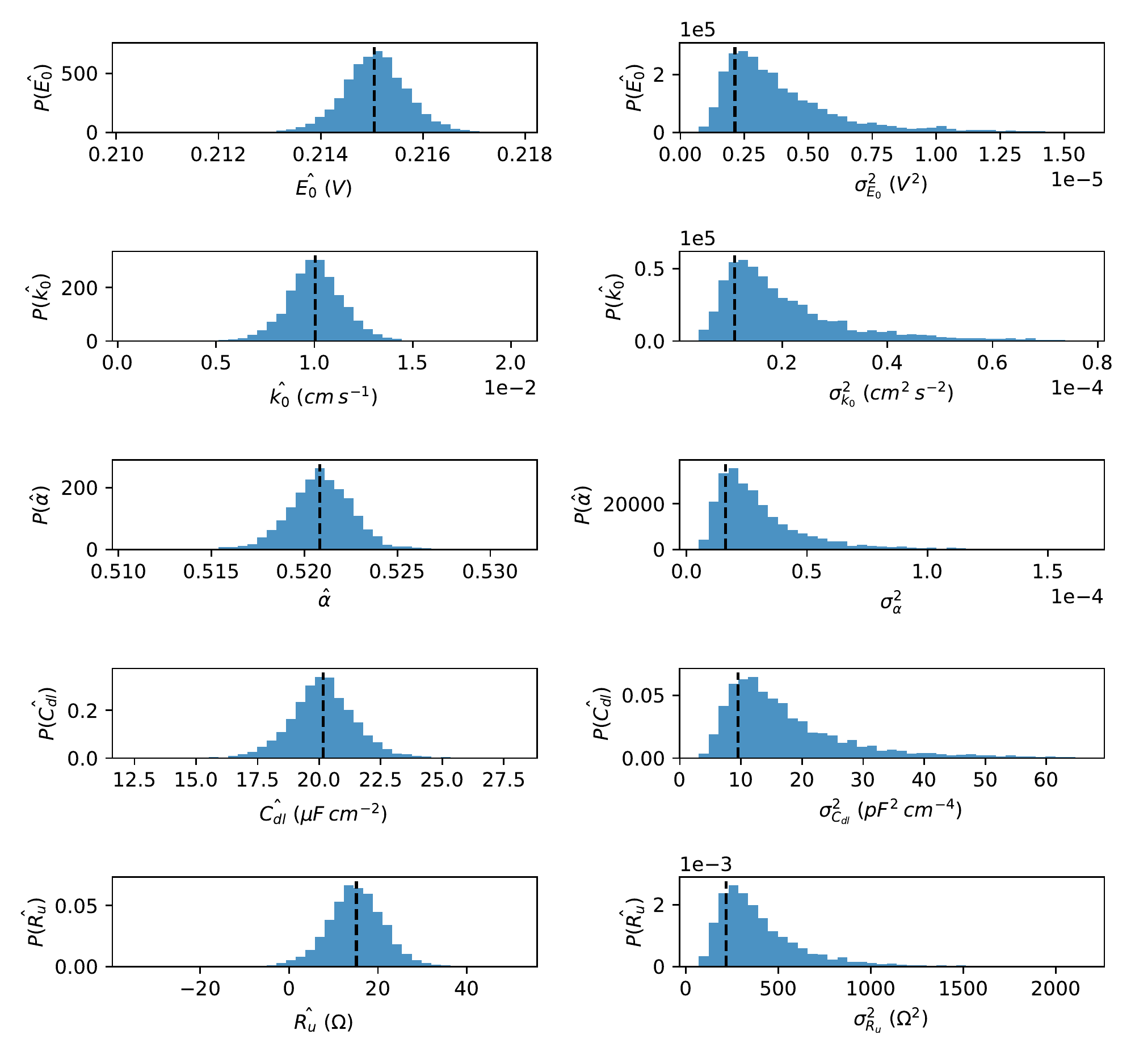}
  \caption{Histograms of hyper-parameter samples of $\boldsymbol\mu$ (left) and
  $\boldsymbol\Sigma$ (right) obtained by applying the hierarchical MCMC
  algorithm \ref{alg:metropolis_within_gibbs} to the ten experimental ac
  voltammetric datasets for the reduction of aqueous 1 mM $[$Fe(CN)$_6]^{3-}$.
  Note that only the diagonal elements of $\boldsymbol\Sigma$ (i.e. the
  variances $\sigma^2$) are shown. For comparison, the dashed lines show the
  sample mean and variance taken across all ten lower level MCMC samples of
  $\boldsymbol\theta_i$.}
  \label{fig:htrace}
\end{figure*}

The chief benefit of the hierarchical model is that it allows us to quantify
(with the hyper-parameters) the variability of the parameters \emph{between}
different experiments. Figure \ref{fig:htrace} shows the histograms for the
hyper-parameters samples of $\boldsymbol\mu$ (left) and $\boldsymbol\Sigma$
(right). Also shown as vertical dashed lines are the the sample mean and
variance of the concatenated ten $\boldsymbol\theta_i$ chains. The most obvious
feature of these plots is that these hyper-parameter distributions clearly show
a much wider confidence interval as compared with the individual
$\boldsymbol\theta_i$ distributions. While the value of each parameter for each
\emph{individual} experiment is known with high accuracy, once the variability
between experiments is taken into account we see that the possible parameter
range is significantly broadened.

\section{Discussion}

The Bayesian data analysis approach introduced in this paper provides access to
fundamentally new knowledge that assists in elucidating nuances that have
contributed to the highly non-reproducible electrode kinetic data published for
the ``pathologically variable" $[$Fe(CN)$_6]^{3-/ 4-}$ process.  Traditional
heuristic
and data analysis optimisation methods produce only single point values for a
limited set of parameters and do not quantify the system variability, which
is crucial information. Using a statistically based Bayesian inference approach,
we are now able to show that the difficulty in achieving reproducibility in the
voltammetry is not associated with the impact of random noise, since for each
data set we are able to fit the experimental data extremely accurately
using a model derived from use of Butler-Volmer electron transfer
kinetics, mass transport by planar diffusion, uncompensated resistance and
double layer capacitance. Thus, while substantial variation in $k_0$ from about
0.002 to 0.018 cm $\text{s}^{-1}$ is evident in 10 individual experiments at a
nominally identical electrode surface (Figure 3), conformance to the
quasi-reversible model is exceptionally good for each individual experiment.

The variation in performance of the now very widely used glassy carbon electrode
was identified as a point of concern soon after the material was introduced into
electroanalytical chemistry \cite{L_van1980glassy}. In essence, the exact nature
of the glassy carbon, and indeed other carbon based surfaces, has their origin
in the method (e.g. temperature) of manufacture, nature of pre-treatment and its
history as an electrode (see for example
\cite{J_kuwana2018analytical,K_kuwana2018analytical,L_van1980glassy,M_declements1996electrochemical,N_dekanski2001glassy,O_ilangovan1997electrochemical,P_declements1996electrochemical,X_chaisiwamongkhol2017amperometric,Y_chaisiwamongkhol2018singlet}
and references cited therein). Accordingly, it seems likely that the
widely variable electrode kinetic data associated with published studies on the
$[$Fe(CN)$_6]^{3-/ 4-}$ process (summarised in Eq.~\ref{eq:chemical_equation})
\begin{equation}\label{eq:chemical_equation}
  \text{Fe(CN)}_6^{3-} + \text{e}^-\cee{<=>}
    \text{Fe(CN)}_6^{4-}\qquad(E_0,k_0,\alpha,C_{dl},R_u),
\end{equation}
mimics the variability of the surface state used in the different publications.

In practice, carbon electrodes are highly heterogeneous with surface defects and
organic functional groups in abundance when used in aqueous electrolyte media
\cite{N_dekanski2001glassy,O_ilangovan1997electrochemical,P_declements1996electrochemical}.
This seems to translate into an electron transfer process occurring at a surface
consisting of microscopically small and distinctly different regions that must
be sufficiently close so that complete overlap of diffusion layers occurs on the
measurement timescale \cite{compton2011understanding}, presumably allowing the
entire surface to be successfully modelled by approximating the mass transfer by
planar diffusion. Thus, even though highly heterogeneous with variable $k_0$
values at the microscopic level, the surface behaves as though it is fully
homogeneous, hence giving rise to measurement of an apparently single "averaged"
$k_0$ value.

Electron transfer reaction mechanisms are often classified into inner and outer
sphere categories in both homogeneous chemical redox reactions that occur in the
solution phase and in heterogeneous reactions in electrochemistry that take
place across an electrode-solution interface\cite{compton2011understanding,BardFaulkner}. In a chemical redox
reaction, the homogeneous outer sphere class mechanism is characterised by weak
interactions of the reactant and product, as in Eq.~\ref{eq:outer_sphere},
\begin{equation}\label{eq:outer_sphere}
  ^*\text{Fe(CN)}_6^{3-}\text{(aq)} + \text{Fe(CN)}_6^{4-}\text{(aq)}\cee{<=>}
  ^*\text{Fe(CN)}_6^{4-}\text{(aq)} + \text{Fe(CN)}_6^{3-}\text{(aq)},
\end{equation}
which is the homogeneous analogue of the electrochemical one of interest in this
study, while in the inner class, ligands involving bridging to a common metal centre may be involved.

In an outer sphere electrochemical process, the plane of closest approach to the Outer Helmholtz Plane does not allow penetration of the layer of non-specifically adsorbed or coordinated solvent adhered to the electrode surface by reactants. In contrast, inner sphere electron transfer processes involve specifically adsorbed reactants \cite{compton2011understanding,BardFaulkner}, and therefore are
anticipated to exhibit electrode kinetics that are strongly dependent on the
chemical nature of the electrode surface. Thus, the $[$Fe(CN)$_6]^{3-/ 4-}$
process can be designated as outer sphere under homogeneous chemical redox
reaction conditions but inner sphere under electrochemical conditions at a
glassy carbon electrode. Presumably, electron transfer at such electrodes is
accompanied by interaction with surface functional groups,
facilitated by the high negative charges associated with the $[$Fe(CN)$_6]^{3-}$
reactant and $[$Fe(CN)$_6]^{4-}$ product which allow specific
binding or electrostatic attractive and repulsive interactions to accompany the
electron transfer process. On this basis, the $[$Fe(CN)$_6]^{3-/ 4-}$ electrode
kinetics are strongly dependent on the treatment and origin of the glassy carbon.

Before attempting to identify the factors that may be most relevant to the
variation encountered in the electrode kinetics, it is worthwhile reviewing the
electrode pre-treatment regime. A nominally 3 mm diameter GC working electrode
was purchased from BAS. Prior to each set of measurements the surface of the GC
electrode was thoroughly polished with 0.3 $\mu \text{m}$ alumina powder in the
form of an aqueous slurry on a wet polishing cloth (BAS). After polishing, the
electrode was repeatedly washed with high purity water and subjected to sonication for 10
to 20 s.  To provide effective removal of any residual alumina powder, after the
initial sonication, the electrode was carefully wiped with a clean wet polishing
cloth, again washed with water and sonicated for 10 to 20 s in high purity water. Finally, the working
electrode
was dried under a nitrogen stream. As far as possible, each electrode
preparation was undertaken under identical conditions; nevertheless, the
electrode kinetics differ substantially from experiment to experiment. However,
of course the history of the electrode could be important since the electrode
used for experiment ten will have a more extensively polished electrode than
that used in experiment one.

The narrow distribution of each $P_i(\cdot)$ in Figure
\ref{fig:bottom_level_chains} enables us to conclude
that the contribution to parameter variability from noise associated with each
particular experiment is minimal for all five parameters estimated. From
perusal of Figures \ref{fig:bottom_level_chains} and \ref{fig:htrace}, we can
conclude that while the degree of experiment-to-experiment
variation changes with respect to each parameter, it is in general significantly
greater than the parameter variability from noise alone (i.e.\ the width of each
$P_i(\cdot)$).  From the
hyper-parameter samples we can
calculate the distribution $P(\cdot)$ of each parameter that would be expected
from another repeat experiment (shown in Figure \ref{fig:bottom_level_chains},
left column). Quantitatively, the parameter mean values (with one standard
deviation values
in parenthesis) as deduced from this analysis are
as follows: $E_0  =  0.215$ V vs
Ag/AgCl (0.002), $k_0 = 0.010$ cm $\text{s}^{-1}$  (0.005) $\alpha = 0.521$
(0.006), $C_{dl} = 20.1$ $\mu F$ $\text{cm}^{-2}$
(4.5), $R_u = 15.2$ $\Omega$ (21.0). Intriguingly, experiment-to-experiment
variation in $\alpha$ is very
small. That is, its value lies in a very narrow range of
about 0.515 to 0.525 with a standard deviation of just 0.006. It is also
notable that while the value of $R_u$ covers a range from zero to about 35
$\Omega$,
it is always small. This means that the more important ohmic drop
term that can distort voltammograms also is small and hence not highly
significant.  The large standard deviation of 21.0 $\Omega$ with a mean value of
15.2 $\Omega$ allows us to
conclude the $R_u$  parameter only has a minor  impact on the ac voltammetry
(highly conducting 3 M \ce{KCl} aqueous electrolyte) and hence on $E_0$, $k_0$,
or $\alpha$
variability. Although the experiment-to-experiment variability of $E_0$ is much
greater than predicted by $P_i(E_0)$, it is still fitted to a tight regime
(about 2 mV) when compared with the total voltage scan range. Indeed, in the
absence of adsorption, $E_0$ is
theoretically predicted to be completely independent of electrode material or
state, suggesting that the inter-experiment variability of $E_0$ is due to
compensating factors from other parameters. $C_{dl}$ displays some variability
with values lying within the range of
about 16 to 23 $\mu F$ $\text{cm}^{-2}$ which confirms that the
electrode is not in an identical surface state, and hence does have a variable
level of activity for each experiment. $k_0$ as noted above encompasses a wide
range of about 0.002 to 0.018 cm $\text{s}^{-1}$ (Figure 3) with a mean value of 0.010 cm $\text{s}^{-1}$ and a
standard deviation of 0.005 cm $\text{s}^{-1}$. The question that arises is
whether there is a correlation between $C_{dl}$ or surface activity and $k_0$.

\begin{figure*}[htbp]
\centering
\includegraphics[width= \textwidth]{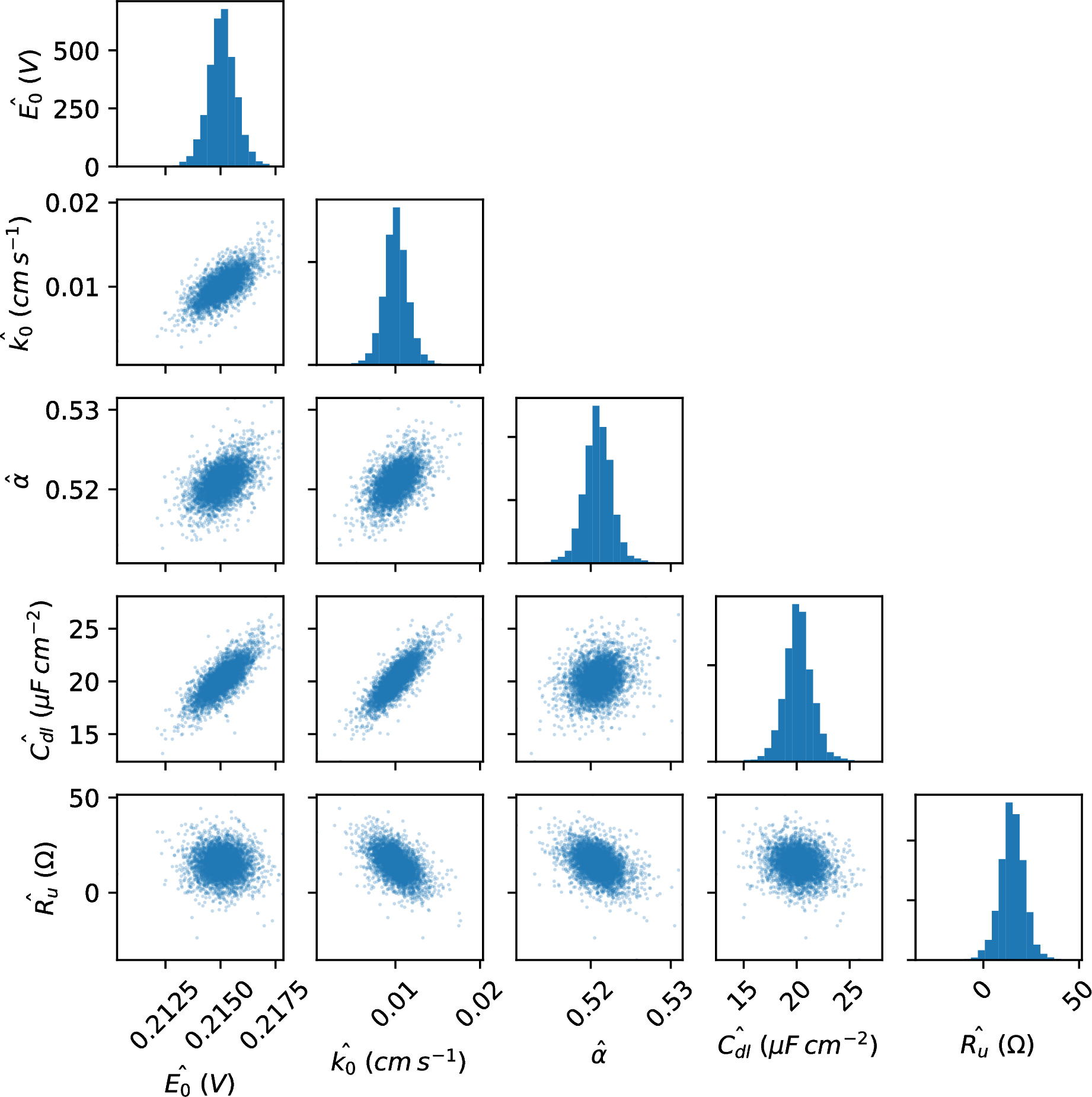}
    \caption{Pairwise correlation plots for the mean hyper-parameter samples
    $\boldsymbol\mu = (\hat{E_0}, \hat{k_0}, \hat{\alpha}, \hat{C_{dl}},
    \hat{R_u})$ (see Algorithm \ref{alg:metropolis_within_gibbs}) obtained for
    10 independent ac voltammetric experiments for the reduction of aqueous 1 mM
    $[$Fe(CN)$_6]^{3-}$. Each dot in the scatter plots shows a sample drawn from
    the posterior distribution $P(\boldsymbol\mu,\boldsymbol\Sigma)$
    (Eq.~\ref{eq:niw}),
    showing the correlation between parameter pairs. The diagonal plots show
    histograms of each individual mean hyper-parameter.}
  \label{fig:pairwise}
\end{figure*}

Figure \ref{fig:pairwise} shows quantitatively the correlation between pairs of
parameters using scatterplots of the mean hyper-parameter $\boldsymbol\mu$
samples.  Clearly, $R_u$ and $\alpha$ are not correlated at all with each other or any other parameter (i.e. they give ``shot
gun" correlation plots). However, there appears to be a weak correlation of $k_0$ and $C_{dl}$ with
larger $k_0$ values being associated with data sets having higher $C_{dl}$ values. There also
appears to be a weak correlation between $k_0$ and $E_0$, with the more positive
$E_0$ values coinciding with the larger $C_{dl}$ values. One needs to be careful
not to over interpret the significance of weak correlations, as other
non-quantified parameters also can be operative. However, a larger capacitance
current caused by a
larger activation of the glassy carbon surface (more surface functionality) may be
expected to increase the rate of an inner sphere process, like the
$[$Fe(CN)$_6]^{3-/ 4-}$  one probed in this work. The origin of the weak
correlation of $k_0$ and $E_0$ is more problematical. The presence of extremely
weak
adsorption not accommodated in the model is one possible explanation, but a
small drift in reference electrode potential cannot be ruled out.

\section{Conclusion}

In summary, the Bayesian inference-inspired strategy introduced in this paper
for data evaluation represents a significant advance in understanding the
contribution of different parameters to voltammetric data and their significance in experiment-to-experiment variability at a heterogeneous electrode in a manner that has not been possible in earlier studies. Perhaps remarkably, each data set in the pathologically variable $[$Fe(CN)$_6]^{3-/ 4-}$
process conforms exceptionally well with simulated data derived from the Butler-Volmer model of electron transfer and mass transport by planar diffusion even though the variation in $k_0$ is quite substantial. Intriguingly, $\alpha$, unlike $k_0$, does not reveal significant experiment-to-experiment variation in this
data analysis exercise. This may be expected if conformance to the Butler-Volmer model is strong.

In early electrode kinetic studies, the ideal and very homogenous mercury
electrode was used. It is now evident that electrode design is becoming very
sophisticated, particularly when advances in materials science are used to
generate highly heterogeneous electrode materials \cite{zhang2018fourier}. Thus,
there is a tendency in electrochemistry nowadays to use far more complex
electrodes than the historically important liquid mercury and pure solid metal
surfaces. The new
breed of highly heterogeneous electrodes imply that data analysis strategies in the future will also need to be more sophisticated. That is, use of models
significantly more complex than used in this study with Butler-Volmer theory for electron transfer and mass transport by planar diffusion, as well as Bayesian
forms of data analysis, will become increasingly essential if informative experimental versus simulation comparisons are to be reported.

\section*{Acknowledgements}

AMB, DJG and JZ would like to acknowledge the support Australian Research Council through the award of a Discovery Grant DP170101535. JZ and ANS also acknowledge the financial support through the ARC Centre of Excellence for Electromaterials Science (ACES). MR gratefully acknowledges research support from the EPSRC Cross-Disciplinary Interface Programme (EP/I017909/1). AMB also wishes to thank the Vallee Foundation for travel support that enabled him to spend time at the University of Oxford.

\bibliography{Alan_paper_refs}

\begin{tocentry}
    \includegraphics[width=0.62\textwidth]{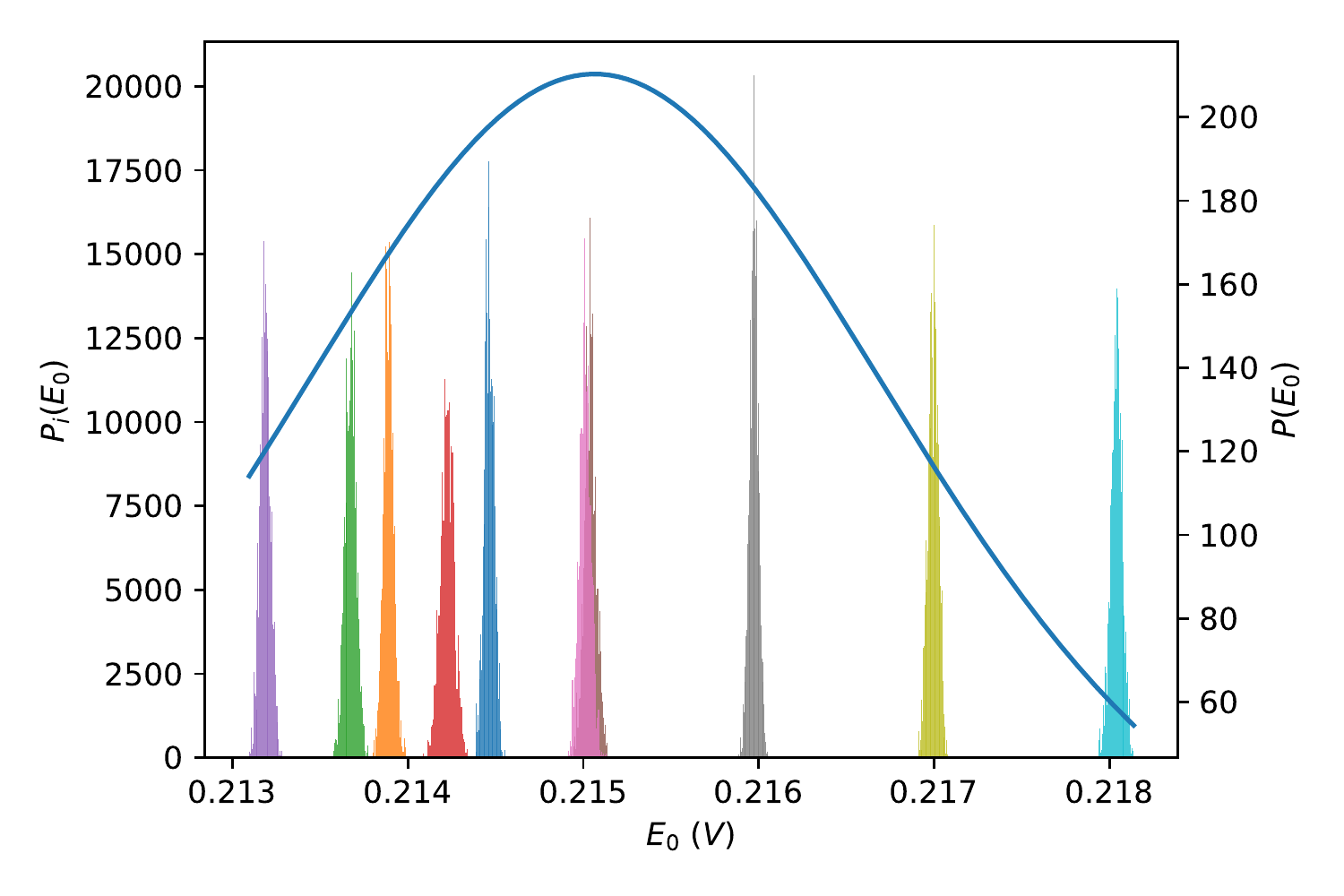}\\
    Distributions for all ten lower level adaptive MCMC samples of
  the reversible formal potential $E_0$. The coloured histograms show the lower
    level posterior for each of the 10 repeat experiments $P_i(E_0)$. The solid
    line shows the predictive posterior for $E_0$, which describes the
    distribution of
    each parameter that would be expected from another repeat of the experiment.
\end{tocentry}

\end{document}